\documentclass[preprint,preprintnumbers,amsmath,amssymb]{revtex4}

\usepackage{epsfig}
\usepackage{tikz}
\usepackage{graphicx} % Include figure files
\usepackage{dcolumn} % Align table columns on decimal point
\usepackage{bm} % bold math
\usepackage{url}
\usepackage{xcolor}

\begin{document}

%\documentclass{article}
%\usepackage[utf8]{inputenc}

%\nofiles

\title{Localizing synergies of hidden factors across complex systems: resting brain networks and HeLa gene expression profile as case studies}

\author{Marlis Ontivero-Ortega$^{1,2}$, Gorana Mijatovic$^3$, Luca Faes$^{4,3}$, Daniele Marinazzo$^{5}$, and Sebastiano Stramaglia$^{1}$}

\affiliation{$^1$ Dipartimento Interateneo di Fisica, Universit\`a degli Studi di Bari Aldo Moro, and INFN, Sezione di Bari, 70126 Bari, Italy\\}

\affiliation{$^{2}$ \quad Cuban Center for Neuroscience, Havana, Cuba\\}

\affiliation{$^{3}$ \quad Faculty of Technical Sciences, University of Novi Sad, 21000 Novi Sad, Serbia\\}
\affiliation{$^4$ Dipartimento di Ingegneria, Universit\`a di Palermo, 90128, Palermo, Italy\\}
\affiliation{$^5$  Department of Data Analysis, Ghent University, Ghent, Belgium\\}

\date{\today}% It is always \today, today,

\begin{abstract} 
Factor analysis is a well-known statistical method to describe the variability of observed variables in terms of a smaller number of unobserved latent variables called factors. Even though latent factors are conceptually independent of each other, their influence on the observed variables is often joint and synergistic. We propose to quantify the synergy of the joint influence of factors on the observed variables using the  O-information, a recently introduced metrics to assess high order dependencies in complex systems, in a new framework where latent factors and observed variables are jointly analyzed in terms of their joint informational character. Two case studies are reported: analyzing resting fMRI data, we find that DMN and FP  networks show the highest synergy, consistently with their crucial role in higher cognitive functions; concerning HeLa cells, we find that the most synergistic gene is STK-12 (AURKB), suggesting that this gene is involved in controlling the HeLa cell cycle. We believe that this  approach, representing a bridge between factor analysis and the field of high-order interactions, will find wide application across several domains.
%\pacs{05.10.-a,05.45.Tp,87.10.-e}
\end{abstract}\maketitle

\maketitle
%\section{Introduction}

\section{Introduction}

Expressing data sets in terms of latent variables offers a powerful way to uncover the hidden structure and underlying drivers in complex systems, making data more manageable and interpretable. 
Factor analysis (FA) \cite{lf} is a powerful statistical method widely used in a variety of fields, including the behavioral sciences \cite{lf_ap_bs}, social sciences \cite{lf_ap_ss}, life sciences \cite{lf_ap_ls}, physical sciences \cite{lf_ap_ps} and business \cite{lf_ap_b}, to describe the variability among observed variables in terms of a smaller number of unobserved latent variables called factors. Specifically, the observed  variables are modeled as linear combinations of the factors, short of some error or noise terms, and this method is extremely effective in reducing the set of relevant variables in large datasets. Factor analysis is closely related to principal component analysis \cite{lf_pca} and therefore to singular value decomposition \cite{lf_svd}. Principal components analysis entails a rotation of variable space in order to capture the maximum variance from the new variables. The three methods become identical when the error terms, or equivalently the variability not explained by the common factors, all have the same variance \cite{lf_bishop}.

It is worth observing that, although latent variables from FA are by construction uncorrelated, they cooperate to produce effects on the observed variables. Therefore it is interesting to localize, in the complex system under consideration, the observed variables whose behavior depends jointly on several factors so as to quantify the synergistic effects from these factors. Intuitively one may expect that such variables, whose role is to integrate the information coming from several factors, are those which depend on more than one factor, as it is is illustrated in figure 1, where two factors influence the observed variables; the observed variables which depend on both factors are  those encoding the joint action of the two factors, thus embodying the cooperation of the two latent variables. The aim of the present work is proposing to quantify the cooperative effects of latent factors  exploiting the notion of synergy which has been recently developed in the framework of partial information decomposition \cite{hoi} to identify the variables responsible of integration in complex systems. In particular, we propose the use of the O-information metrics \cite{hoi_oinf} to quantify cooperation and localize synergistic effects in the system. We like to mention that another approach to localize high order effects in complex systems has been proposed in \cite{hoi_grad} using gradients of the O-information evaluated on groups of observed variables: the peculiarity of the method, here introduced, is that we introduce latent variables to discard redundancy, so that the net effect of latent variables on the observed variables can only be synergistic.

\section{Methods}

Let us recall the definition of O-information \cite{hoi_oinf}, a metric measuring the balance between redundancy and synergy, the two basic types of high order statistical dependencies. Let ${\bf x} =\{x_1,\ldots,x_n\}$ denote a set on $n$ stochastic variables, and  let $\bf{x}_{-i}$ denote the set of all the variables in $\bf{x}$ but $x_i$. 
The O-information is defined as \cite{hoi_oinf}:
%First of all, we recall that  O-information $\Omega$ is a signed metric that measures the balance between redundant and synergistic interactions within a system described by $n$ stochastic variables $\bm{X}^n=(X_1,\dots,X_n)$, which is evaluated as follows~(\cite{rosas_quantifying_2019}:
\begin{equation}
    O_{inf}({\bf {x}}) =  
         (n-2)H({\bf x}) + \sum_{i=1}^n \Big[ H(x_i) - H({\bf x_{-i}}) \Big],
         \label{Oinfo}
\end{equation}
where H is the Shannon entropy.
If $O_{inf}>0$, the system is redundancy-dominated. On the other hand, when $O_{inf}<0$ the dependencies are better explained as patterns that can be observed in the joint state of multiple variables but not in subsets of these; in other words, the system is synergy-dominated.

Now, we recall the generative model for FA, i.e.:
$${\bf x}= {\bf L \;f}+ \underline{\eta},$$
where ${\bf x}$ is the $N$-dimensional vector of observed variables, ${\bf f}$ is a vector of $m$ latent factors, ${\bf L}$ is the $N\times m$ matrix of loadings whilst $\underline{\eta}$ is a vector of $N$ Gaussian noise terms with variances $\underline{\sigma}$. Factors are assumed to be Gaussian zero mean unit variance independent variables,  factors and noise terms being statistically independent.
Given a suitable number of samples of ${\bf x}$, both loadings and factor scores can be estimated by maximum likelihood.

For clarity, we concentrate now on the case of a group of latent factors influencing a single target variable.
Let $f_1, f_2, \ldots, f_{n-1}$ be $n-1$ independent  latent variables acting as drivers for the target $$x=\sum_{i=1}^{n-1} L_i f_i +\eta,$$ where $\eta$ is a Gaussian noise term with variance $\sigma^2$. The structure corresponding to this equation is a collider, corresponding to net synergy. Without loss of generality, we assume that all $f's$ are zero mean unit variance Gaussian variables, and that $x$ has unit variance, i.e. $\sum_{i=1}^{n-1} L_i^2 +\sigma^2=1.$
Straightforward calculations, based on the formula for the entropy of a multivariate Gaussian distribution, provide the O-information for the group of $n$ variables $\{x,f_1,\ldots,f_{n-1}\}$:

\begin{equation}
O_{inf}={1\over 2}\log{\left(1-\sum_{i=1}^{n-1} L_i^2\right)^{n-2} \over \prod_{i=1}^{n-1} \left(1-\sum_{k\ne i} L_k^2\right)}
\end{equation}
%For small $L's$, $O$ can be approximated as $$O\sim -{1\over 2} \left(n-2\right)^2 \sum_{i\ne k} L_i^2L_k^2.$$

Note that this quantity is always less than zero or it vanishes, thus confirming that the occurrence of net synergy. We remark that $O_{inf}$ is zero if all $L's$ vanish but one. For small $L's$, $O_{inf}$ is order $$O_{inf}\sim - \sum_{i\ne k} L_i^2L_k^2.$$ The formula above quantitatively supports the scenario described in figure 1, i.e. an observed variable is synergistic for two latent factors only if it is dependent on both latent factors.

The calculation above suggests the following procedure to highlight synergy: (i) fix the number of latent factor, (ii) fit the FA model to data and extract the latent factor scores, (iii) for each observed variable, evaluate $O_{inf}$ for the multiplet constituted by the observed variable plus the latent factors, $-O_{inf}$ representing the net synergy; (iv) variables corresponding to low negative values of $O_{inf}$ are those responsible for the integration in the system.

We remark that to evaluate $-O_{inf}$ one may either take the estimated loading matrices and  use eq. (2), or  perform a direct evaluation  using the estimated samples of factor scores and the measured variables, as we did in this work.

\begin{figure}[!ht] \centering
\includegraphics[width=.65\textwidth]{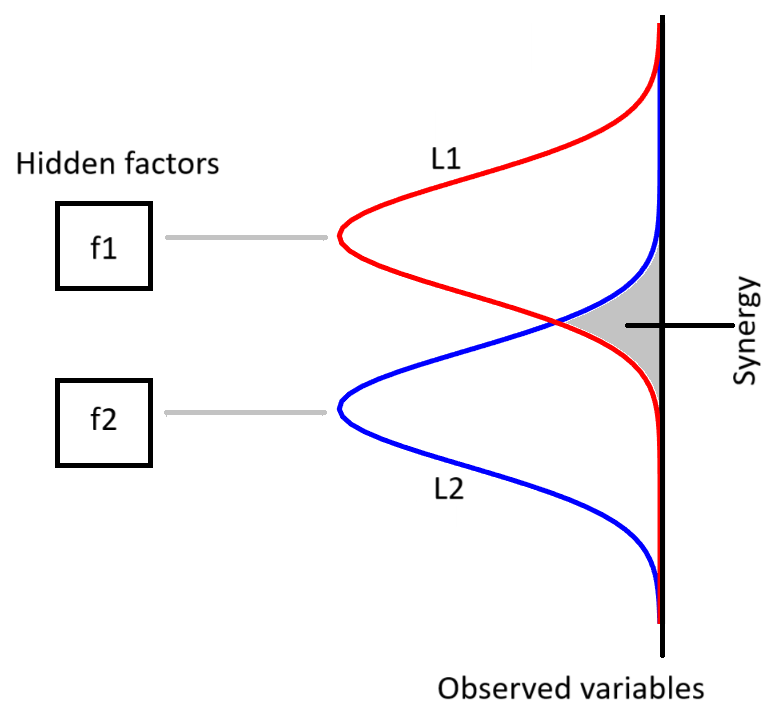}
\caption{Cartoon representing intuitively the proposed approach. Two hidden factors influencing the observed variables: the overlap between the two loading profiles $L_1$ and $L_2$ corresponds to variables which are synergistic for factors $f_1$ and $f_2$. }
\label{fig:atm}
\end{figure}

\section{Results}

In the following, we describe two applications of the proposed framework in biomedical engineering, i.e. the analysis of resting state fMRI time series of healthy subjects, and the analysis of gene expression data from HeLa cells.

\subsection{fMRI data}

We used the public dataset described in Poldrack et al. \cite{fmri_data}. These data was obtained from the OpenfMRI database, with accession number ds000030, and were already used in \cite{fmri_vis}. We use resting-state fMRI data from 121 healthy controls. The demographics are reported in the original paper.
%, and they can additionally be found in \cite{git}.

Data were preprocessed with FSL (FMRIB Software Library v5.0) \cite{fmri_fsl}. The volumes were corrected for motion, after which slice timing correction was applied to correct for temporal alignment. All voxels were spatially smoothed with a 6 mm FWHM (full width at half maximum) isotropic Gaussian kernel and after intensity normalization, a band pass filter was applied between 0.01 and 0.08 Hz. In addition, linear and quadratic trends were removed. We next regressed out the motion time courses, the average cerebrospinal fluid signal, and the average white matter signal. Global signal regression was not performed. Data were transformed to the MNI152 template, such that a given voxel had a volume of 3 mm × 3 mm × 3mm. Finally, we averaged the signal in 268 ROIs. In order to localize the results within the intrinsic connectivity network of the resting brain, we assigned each of these ROIs to one of the nine resting-state networks (seven cortical networks, plus subcortical regions and cerebellum) as described in \cite{fmri_atlas}.
Time series from subjects were firstly z-scored and then concatenated in a matrix with  $268$ observed variables and $152\times 121=18392$ samples.

We applied the factor analysis function {\it factoran} of MATLAB to fit these four matrices, with the number of hidden factors fixed at 20 (we found that results are robust to slight variations of the number of factors). Some factors thus obtained were recognized to be related to  common trend of data: therefore we removed from the subsequent analysis three factors mostly correlated to the common trend,
and we were left with 17 factors for each of the four groups of subjects.

Then, for each region, we evaluated the O-information of that region and the 17 latent factors; the statistical significance of $O_{inf}$ can be assessed using surrogates obtained by random permutation of the samples of the target region. 

The results are shown in figure 2.
We find that FP and DMN show the highest synergy, and that generically synergy is higher in associative areas supporting cognitive functions and lower in sensory-motor areas, consistently with previous analyses which use other tools to evaluate higher order dependencies \cite{fmri_luppi, fmri_scag,fmri_varley}.

\begin{figure}[!ht] \centering
\includegraphics[width=.8\textwidth]{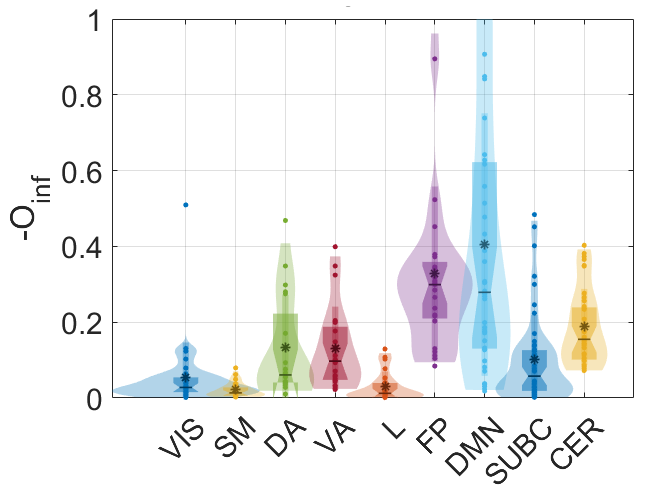}
\caption{The opposite of the O-information (measuring net synergy) is plotted for the 268 regions in the fMRI dataset.}
\label{fig:atm}
\end{figure}

\subsection{HeLa data}

We apply the proposed approach to data from the cell culture HeLa. The data correspond to 94 genes and 48 time-points \cite{hela_data},
with an hour interval separating two successive readings (the HeLa cell cycle lasts 16h). The 94 genes were selected, from
the full dataset described in \cite{hela_dataFull}, on the basis of the association with cell cycle regulation and tumor development.
Since the number of samples is less than the number of variables, in this case we use principal component analysis instead of FA.
As described in \cite{hela_data1}, the first two principal components show exponentially decaying correlations whilst the third principal component seems to be connected with cell cycle as it shows oscillations with a period close to 16. 
We discard the first principal component and consider here the second and the third components to find the genes which exhibit synergy w.r.t. these two factors.

The results are depicted in figure 3; 24 genes, out of 94, show a synergy which is statistically higher than those from surrogates after Bonferroni correction. The highest synergy is obtained in correspondence of STK-12 (Aurora Kinase B), which is known  to function as a transcriptional brake, controlling the expression of genes involved in cellulase production \cite{hela_stk}.
A synergistic interaction between Aurora B and ZAK has been found  in triple-negative breast cancer \cite{hela_stk_canc}: our analysis shows that it may play a key role also in the control of HeLa cell cycle.

\begin{figure}[!ht] \centering
\includegraphics[width=1\textwidth]{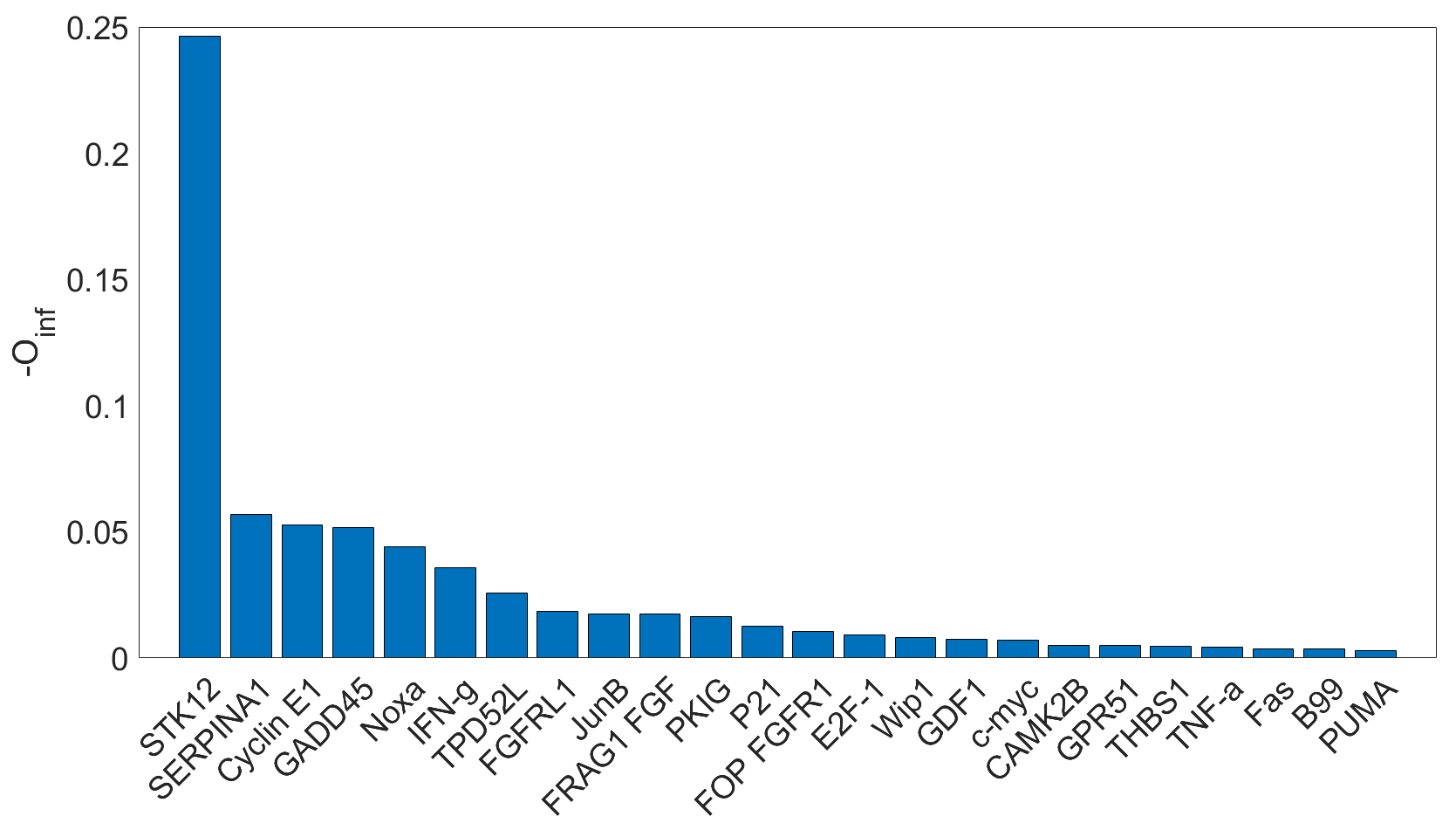}
\caption{The opposite of the O-information (measuring net synergy) is plotted for the 24 genes in the HeLa data-set whose synergy is recognized as statistically significant against surrogates after Bonferroni correction.  The highest synergy is obtained in correspondence of the gene STK12.}
\label{fig:atm}
\end{figure}

\section{Discussion}

Many complex systems can be described in terms of latent factors which influence the observed variables. The description in terms of latent factors can be seen as a way to get rid of the redundancy in data, hence it can be exploited to highlight synergies in data sets where synergies are obscured by redundancy.
In this work, we have proposed a new perspective on high order dependencies analysis, i.e. the evaluation of the informational character of circuits involving both latent factors and the observed variables: this allows to localize in the system the variables whose behavior depends synergistically on factors and therefore play the role of integrators of information.
In particular, we propose to evaluate the informational character of the group of variables made by factors and each given observed variable using the O-information, thus quantifying the synergistic role played by each observed variable. 

We apply the methodology to fMRI data from  healthy individuals, and found less synergy in sensory networks respect to networks which support complex cognitive processes such as planning and execution of goal-directed behavior \cite{fp_net}. %different synergistic patterns for schizophrenia in cerebellum and visual networks. %and subcortical resting networks are less involved in synergies than the other resting networks.%  
Applying our method on genetic data from HeLa cells, we found that the most synergistic gene is STK-12, whose synergistic role was already assessed performing the analysis at the level of observed gene expressions. 

We believe that the proposed approach may add further insights in those complex systems which allow a suitable representation in terms of latent variables. 
Further research will be devoted to implement the proposed approach with other approaches for the inference of latent factors, as well as to focus on the problem of selecting the proper factors to include in the analysis. Another interesting issue will be to study the relation between the synergies of factors (here introduced) and those that can be measured on groups of observed variables, i.e. the relation between mechanisms and behaviors in this context \cite{hoi_rosas_mecbeh}.

%\authorcontributions{methodology, MOO and SS; software, MOO and SS;  writing---original draft preparation, MOO, GM, LF, DM, and SS; writing---review and editing, MOO, GM, LF, DM, and SS. All authors have read and agreed to the published version of the manuscript.}
\begin{acknowledgments}
This research was funded by the project “HONEST - High-Order Dynamical Networks in Computational Neuroscience and Physiology: an Information-Theoretic Framework”, Italian Ministry of University and Research (funded by MUR, PRIN 2022, code 2022YMHNPY, CUP: B53D23003020006) (MOO, LF and SS); and by the project “Higher-order complex systems modeling for personalized medicine”, Italian Ministry of University and Research (funded by MUR, PRIN 2022-PNRR, code P2022JAYMH, CUP: H53D23009130001) (SS).
\end{acknowledgments}
%\acknowledgments{In this section you can acknowledge any support given which is not covered by the author contribution or funding sections. This may include administrative and technical support, or donations in kind (e.g., materials used for experiments). Where GenAI has been used for purposes such as generating text, data, or graphics, or for study design, data collection, analysis, or interpretation of data, please add “During the preparation of this manuscript/study, the author(s) used [tool name, version information] for the purposes of [description of use]. The authors have reviewed and edited the output and take full responsibility for the content of this publication.”}

%\conflictsofinterest{The authors declare no conflicts of interest.} 

%\reftitle{References}
%\isAPAandChicago{}{%


\begin{thebibliography}{999}

%\bibitem{xai} S. Ali, T. Abuhmed, S. El-Sappagh, K. Muhammad, J.M. Alonso-Moral, R. Confalonieri, R. Guidotti, J. Del Ser, N. Díaz-Rodríguez, and F. Herrera, \emph{Information Fusion} {\bf 99}, 1566 (2023).

\bibitem{lf} S. Everitt. An Introduction to Latent Variable Model, Chapman and Hall, London, 1984.

\bibitem{lf_ap_bs} J. B. Carrol. Human Cognitive Abilities: A Survey of Factor-analytic Studies, Cambridge University Press, Cambridge, 1993.

\bibitem{lf_ap_ss} J. P. Stevens. Applied Multivariate Statistics for the Social Sciences, Psychology Press, London, 2002.

\bibitem{lf_ap_ls} H. H. Harman. Modern Factor Analysis, The University of Chicago Press, Chicago, 1976.

\bibitem{lf_ap_ps} D. Love, D. K. Hallbauer, A. Amos, R. K. Hranova. Physics and Chemistry of the Earth 29 (2004), 1135.

\bibitem{lf_ap_b} D. W. Stewart. Journal of Marketing Research 18 (1981) 51.

\bibitem{lf_pca} R. O. Duda, P. E. Hart, D. G. Stork. Pattern Classification, John Wiley $\&$ Sons, New York, 2001.

\bibitem{lf_svd} M. E. Wall, A. Rechtsteiner, L. Rocha. Singular value decomposition  and principal component analysis, in: D.Berrar, W.Dubitzky, M.Granzow(Eds.), A Practical Approach to Microarray Data Analysis, Kluwer Academic Publishers, 2003, pp.91–109.

\bibitem{lf_bishop} M. E. Tipping, C. M. Bishop. Journal of the Royal Statistical Society. Series B 21(1999), 611.

\bibitem{hoi} Williams, P. L., and Beer, R. D. (2010). Nonnegative decomposition of multivariate information. ArXiv.

\bibitem{hoi_oinf} Rosas, F. E., Mediano, P. A. M., Gastpar, M., and Jensen, H. J. (2019). Quantifying high-order interdependencies via multivariate extensions of the mutual information. Phys. Rev. E 100:032305.

\bibitem{hoi_grad} T. Scagliarini et al. Phys. Rev. Research 5, 013025 (2023)

\bibitem{fmri_data} Poldrack, R. A., et al (2016). A phenome-wide examination of neural and cognitive function. Scientific data, 3(1), 1-12.

\bibitem{fmri_vis} S. Sannino, S. Stramaglia, L. Lacasa, D. Marinazzo, Network Neuroscience (2017) 1 (3): 208–221.

%\bibitem{git} https://github.com/danielemarinazzo/Visibility\LA5C\data

\bibitem{fmri_fsl} M. Jenkinson, C.F. Beckmann, T.E. Behrens, M.W. Woolrich, S.M. Smith. FSL. NeuroImage, 62:782-90, 2012 

\bibitem{fmri_atlas} Yeo, B. T., Krienen, F. M., Sepulcre, J., Sabuncu, M. R., Lashkari, D., Hollinshead, M., et al (2011). The organization of the human cerebral cortex estimated by intrinsic functional connectivity. Journal of Neurophysiology, 106(3), 1125–1165. 

\bibitem{fmri_luppi} A. Luppi et al, Nat Neurosci . 2022 June 01; 25(6): 771–782

\bibitem{fmri_scag} Scagliarini T, Sparacino L, Faes L, Marinazzo D and Stramaglia S (2024), Gradients of O-information highlight synergy and redundancy in physiological applications. Front. Netw. Physiol. 3:1335808.

\bibitem{fmri_varley} Varley, T. F., Pope, M., Faskowitz, J. et al. Multivariate information theory uncovers synergistic subsystems of the human cerebral cortex. Commun Biol 6, 451 (2023).

\bibitem{hela_data} M. L. Whitfield, et al., Molecular Biology of the Cell 13 (2002) 1977.

\bibitem{hela_dataFull} A. R. Hoerl, R. W. Kennard. Technometrics 12 (1970) 55.

\bibitem{hela_data1} M. Zamparo, S. Stramaglia, J.R. Banavar, and A. Maritan, Physica A 391 (2012) 3159–3169. 

\bibitem{hela_stk} Lin L., Wang S., Li X., He Q., Benz J. P., Tian C. (2019) STK-12 acts as a transcriptional brake to control the expression of cellulase-encoding genes in Neurospora crassa. PLoS Genet 15(11): e1008510.

\bibitem{hela_stk_canc} Tang, J., Gautam, P., Gupta, A. et al. Network pharmacology modeling identifies synergistic Aurora B and ZAK interaction in triple-negative breast cancer. npj Syst Biol Appl 5, 20 (2019).

\bibitem{fp_net} Hwang, E. J., Sato, T. R., and Sato, T. K. (2021). A canonical scheme of bottom-up and top-down information flows in the frontoparietal network. Frontiers in neural circuits, 15, 691314.

\bibitem{hoi_rosas_mecbeh} F. Rosas et al., NATURE PHYSICS. 2022;18(5):476–7.

\end{thebibliography}
\end{document}